\documentclass[twocolumn,aps,showpacs,floats,letterpaper,floatfix,groupedaddress,eqsecnum]{revtex4}
\usepackage{ulem,amssymb,amsmath}
\usepackage[dvips]{graphicx}
\usepackage{dcolumn,epsfig}

\newcommand{\Tcal}{{\cal T}}

\newcommand{\beq}{\begin{equation}}
\newcommand{\eeq}{\end{equation}}
\newcommand{\bes}{\begin{subequations}}
\newcommand{\ees}{\end{subequations}}
\newcommand{\bea}{\begin{eqnarray}}
\newcommand{\eea}{\end{eqnarray}}
\newcommand{\ba}{\begin{array}}
\newcommand{\ea}{\end{array}}
\newcommand{\beqn}{\begin{eqnarray*}}
\newcommand{\eeqn}{\end{eqnarray*}}

\newcommand{\f}[2]{\frac{#1}{#2}}

\newcommand{\la}{\langle}
\newcommand{\ra}{\rangle}

\newlength{\sizeonefig}
\newlength{\sizetwofig}
\begin{document}

\title{Viscous corrections to the resistance of nano-junctions:\\ a dispersion relation approach}
\author{Dibyendu Roy$^1$, Giovanni Vignale$^2$ and  Massimiliano Di Ventra$^1$}
\affiliation{$^1$Department of Physics, University of California-San Diego, La Jolla, CA 92093. \\$^2$Department of Physics and Astronomy, University of Missouri, Columbia, Missouri 65211, USA.}

\begin{abstract}
It is well known that the viscosity of a homogeneous electron liquid diverges in the limits of zero frequency and zero temperature. A nanojunction breaks translational invariance and necessarily cuts off
this divergence. However, the estimate of the ensuing viscosity is far from trivial. Here, we propose an approach based on a Kramers-Kr\"onig dispersion relation, which connects the zero-frequency viscosity, $\eta(0)$,
to the high-frequency shear modulus, $\mu_{\infty}$, of the electron liquid via $\eta(0) =\mu_{\infty} \tau$, with $\tau$ the junction-specific momentum relaxation time. By making use of a simple formula derived from time-dependent current-density functional theory we then estimate the many-body contributions to the resistance for an integrable junction potential and find that these viscous effects may be much larger than previously suggested for junctions of low conductance.
\end{abstract}

\maketitle
\section{Introduction}
Viscosity, namely the effect of momentum transfer between adjacent layers of a liquid, is a fundamental
concept in both classical and quantum physics \cite{Pines, Landau87, Giuliani05}.  In the case of the electron liquid, it was shown by Abrikosov and Khalatnikov (AK) more than fifty years ago that the viscosity of a homogeneous liquid diverges at zero frequency and zero temperature (in this precise order of limits)\cite{Abrikosov59}. The physical reason for this divergence is related to the fact that at zero temperature the quasi-particles close to
the Fermi energy are very long-lived and can transport momentum to distances arbitrarily far away from the source of the perturbation. We thus expect that any potential that breaks translational invariance would cut off
such divergence giving rise to a finite (albeit possibly large) d.c. viscosity at zero temperature.

The problem has garnered renewed attention in the past few years in the context of
transport in nanoscale systems~\cite{Diventra08}, where it was shown explicitly that the Landauer formula for the single-particle resistance of a nanojunction $R_{\rm s}=h/(2e^2\sum_j {\mathcal T}_j)$
(the sum is over the eigenchannels of transmission ${\mathcal T}_j$)~\cite{Butt1}, fails to include certain many-body effects, which cannot be described as single-particle scattering from an effective potential.~\cite{Sai05,Sai07} Within the framework of the time-dependent current density functional theory~\cite{VK,VUC}, one clearly sees that such effects arise from the frequency dependence of the exchange-correlation (xc) field.~\cite{Vignale09} More precisely, one can split the full exchange-correlation potential into a static component, which controls the transmission probability $\Tcal_j$ of the Landauer formula, and a dynamical component, which corrects the Landauer formula~\cite{Vignale09}. Indeed, even though the control of nanoscale junctions at the atomic level is far from being ideal, it is now clear that the Landauer formula computed within the framework of ground-state
density-functional theory (DFT), overestimates the measured conductance by at least an
order of magnitude in the case of low-conductance structures (e.g., molecular structures), while it provides reasonable agreement in the case of metallic quantum point contacts, which show high-conductance values~\cite{MDlong}. Part of this discrepancy has been attributed to errors in determining the position of the
energy levels of the system relative to the electro-chemical
potential in the leads -- errors which in turn are related to self-interaction corrections, discontinuities in the
xc potential as a function of particle number, and so
on.~\cite{Burke} However, these corrections do not fully solve the discrepancy between theory and experiments: other many-body effects, in particular those related to the viscous nature
of the electron liquid, may play an important role.

\section{Dynamical corrections to the resistance}
Time-dependent current density functional theory provides some insight into the physical character of these many-body corrections. Indeed, it was shown that the dynamical corrections to the xc field give rise to a viscous force~\cite{VK,VUC}, similar to the ordinary hydrodynamic viscous force, but controlled entirely by electron-electron interactions - the ordinary hydrodynamic viscosity does not explicitly depend on interactions, but relies implicitly on the presence of collisions capable of establishing a local thermodynamic equilibrium~\cite{Landau87}. Thus, the resistance in excess of the Landauer formula could be interpreted as the effect of the extra ``friction" arising from the xc viscosity.
In Ref.~\cite{Sai05}, this additional many-body resistance - which was termed ``dynamical'' precisely because it vanishes in a strictly ground-state formulation of the theory~\cite{Vignale09} - was estimated (assuming no
current density variation in the junction) in terms of the exchange-correlation (xc) viscosity of the liquid~\cite{Sai05,Sai07}
\bea
R_{\rm dyn}=\f{\eta_{xc}(0)}{e^2S^2}\int\left[\f{4}{3}(\partial_xn^{-1})^2+(\partial_{\perp}n^{-1})^2\right]d\vec{r}\,,
\label{dynres}
\eea
where $x$ is the direction of current flow through the junction, $\perp$ represents the transverse directions, $S$ is the cross-section area of the nano-junction, $n$ is the ground-state electron density calculated, e.g., from the self-consistent static density functional theory (DFT) with the xc functional treated within the local-density approximation (LDA), $\partial_x$ and $\partial_\perp$ are derivatives along the current flow and perpendicular to it, respectively. Here, $\eta_{xc}(0)>0$ is the zero-frequency xc viscosity of the {\it homogeneous} electron liquid at the average electron density of the junction, and $e$ is the charge of the electron. Thus the dynamical correction increases the total resistance of nano-junctions to $R_{\rm tot}=R_{\rm s}+R_{\rm dyn}$~\cite{Note1}. Notice that the non-uniformity of the electron density is essential to the effect:  $R_{\rm dyn}$ vanishes, as it should, if $n$ is spatially uniform.

The precise definition of $\eta_{xc}(0)$ in current density functional theory is \cite{Conti99}
\bea\label{defetaxc}
\eta_{xc}(0)=  -n^2 \lim_{\omega \to 0} \frac{\Im m f_{xc,T}(\omega)}{\omega}\,,
\eea
where $f_{xc,T}(\omega)$ is the transverse component of the dynamical exchange-correlation kernel of the homogeneous electron liquid at density $n$.  The kernel itself is defined as the difference between the inverse current-current response functions of the interacting and non-interacting system at the same density.  This quantity admits a perturbative expansion in the interaction parameter $r_s$ of the electron liquid -- the average distance between electrons expressed in units of the Bohr radius.  Our initial estimates of $R_{\rm dyn}$ were based on an extrapolated high-density expansion of $\eta_{xc}$, which gave $R_{\rm dyn}$ of about $10\%$ of $R_s$ for molecular
junctions, but considerably smaller for metallic quantum point contacts~\cite{Sai05,Sai07}.  Since then, the use of a more accurate expression for $\eta_{xc}$ as a function of $r_s$, namely \cite{Conti99}
\bea\label{eta_infinity}
\eta_{xc} \simeq \f{\hbar n}{60r_{s}^{-3/2}+80r_{s}^{-1}-40r_{s}^{-2/3}+62r_{s}^{-1/3}}\,,
\eea
has been shown to produce considerably smaller corrections for the case of two infinite jellium electrodes separated by a vacuum
gap.~\cite{Jung07} Indeed, this $\eta_{xc}$ is of the order of $10^{-7}$ Joule-sec/m$^3$ for typical metallic densities such as $r_{s}=3$ for gold, thus naively suggesting that these dynamical corrections are small
under all circumstances.

Before jumping to conclusions, however, it must be noted that the expression~(\ref{eta_infinity}) is not the appropriate zero-frequency viscosity, as required by Eq.~(\ref{defetaxc}), but rather was calculated in Ref.~\cite{Conti99} under the implicit assumption $\omega\gg 1/\tau$ where $\tau$ is the momentum relaxation time for a quasi-particle.   This condition is easily satisfied in the homogeneous electron liquid in the limit of zero temperature $T$, since $\tau$ (in the absence of impurities) tends to infinity as $1/T^2$. It is certainly not satisfied in a nanojunction where the main limiting factor to the quasi-particle lifetime is the geometric size of the constriction itself.~\cite{Diventra08,MD04,DD,DD1} Indeed, it is precisely the geometrical constriction experienced by the electron wave packets as they move into the nanojunction that introduces a short momentum relaxation time~\cite{Diventra08,MD04}, which in turn cuts off the divergence of the viscosity of the uniform electron liquid.

We are then led to consider the opposite and physically correct limit of $\omega \ll 1/\tau$.  But here, we run into the problem that the non-uniformity of the electron liquid must be fully taken into account.  If, for instance, one calculates the d.c. viscosity of the {\it homogeneous} electron liquid, using the techniques developed by AK \cite{Abrikosov59} one obtains  (employing the Thomas-Fermi approximation for the screened electron-electron (e-e)
interaction)
\bea
\eta_{AK}=\f{\hbar n}{(\tilde{r}_s)^6}\big(\f{1.813 \times 10^3}{T}\big)^2\Big \{\f{\pi(1+2\tilde{r}_s)}{8\sqrt{\tilde{r}_s+\tilde{r}_s^2}}-\f{\pi}{4}\Big\}^{-1}\,, \label{viscosityAK}
\eea
where $\tilde{r}_s=\alpha r_s/\pi$ and $\alpha=(4/9\pi)^{1/3} \simeq 0.521$.
This expression has two major shortcomings.  First and foremost, it diverges at low temperature as $1/T^2$, which is the consequence of undisturbed momentum transport by the long-lived quasi-particles of the homogeneous electron liquid to distances arbitrarily far away from the source of perturbation.  Even at room temperature $\eta_{AK}=1.12 \times 10^4/T^2$ Joule-sec/m$^3$ (temperature T is in Kelvin) for gold ($r_s=3$), which, if used in Eq.~(\ref{dynres}) would produce unreasonably large values of $R_{\rm dyn}$.   Second, the AK viscosity is conceptually different from the xc viscosity: it arises from the full current-current response function, not just from the exchange-correlation kernel, and there is no simple way to separate the latter, since the AK calculation is non-perturbative with respect to the electron-electron coupling strength.  Instead, in our case, such a separation is essential, since the larger (elastic) part of the resistance is already taken into account exactly by the Landauer formula, and only the xc kernel contributes to the dynamical (inelastic) correction. Thus, we need a way to ``zero-in" on the xc viscosity (defined formally in terms of the xc kernel $f_{xc}$ in Eq.~(\ref{defetaxc})) just as we did in Ref.~\cite{Conti99} --  but now we must go to the opposite regime of $\omega\ll 1/\tau$.


 Equipped with this understanding, we first show that the viscosity of the electron liquid {\it in the nanojunction} can be estimated by a dispersion relation approach, which combines information about the high-frequency elastic properties of the interacting electron liquid with the short momentum relaxation time induced by the constriction.  We then estimate the dynamical contribution to the resistance of nanostructures from Eq.~(\ref{dynres}) and  show that this contribution is relatively small for nearly transparent junctions but becomes sizeable in the limit of zero transmission.

\section{Dispersion relation approach}

Let us then start by recalling that the viscosity $\eta(\omega)$ and the shear modulus $\mu(\omega)$, regarded as functions of frequency, are respectively the imaginary and the real part of a visco-elastic modulus $\tilde \mu(\omega)\equiv \mu(\omega)-i\omega \eta(\omega)$.~\cite{Landau87, Giuliani05, Conti99}  The shear modulus vanishes at zero frequency, since we are in a liquid state, and tends to a finite value $\mu_{\infty}$ at infinite frequency, where by ``infinite'' we mean a frequency much larger than $1/\tau$.  On the other hand, the viscosity, $\eta(\omega)$, varies from the desired value, $\eta(0)$, at zero frequency to zero at ``infinite'' frequency.
Then, the Kramers-Kr\"onig dispersion relation tells us that
\bea
0=\mu(0)=\mu_\infty -\frac{2}{\pi}\int_0^\infty\eta(\omega)d\omega\,.\label{maxwell1}
\eea
We do not know the detailed frequency dependence of $\eta(\omega)$, but we can assume that it roughly switches from the d.c. value $\eta(0)$ for $\omega<1/\tau$ to approximately zero for $\omega>1/\tau$.
This simple reasoning leads from Eq.~(\ref{maxwell1}) to the relation~\cite{Diventra08,DD1}
\bea \label{Maxwell}
\eta(0)\simeq \mu_{\infty}\tau\,.
\eea
The importance of this relation, in the present context, is that it connects the quantity of primary interest, the zero-frequency viscosity,  to two quantities that can be rather easily estimated, namely, the high-frequency shear modulus -- a positive definite quantity that can be expressed in terms of the exact energy of the electron liquid -- and the momentum relaxation time $\tau$, which, as we have discussed above, in a nano-junction  is mostly controlled by elastic boundary scattering from the confining geometry.  A crucial feature of this relation is that it allows a neat separation of $\eta(0)$ into two parts:  a ``single-particle'' contribution, which arises from
the non-interacting kinetic part of the shear modulus (see Eq.~(\ref{muxc}) below), and an ``exchange-correlation" contribution,  which actually determines, via Eq.~(\ref{dynres}), the dynamical correction to the resistivity.

\begin{figure}[t]
\begin{center}
\includegraphics[width=7cm]{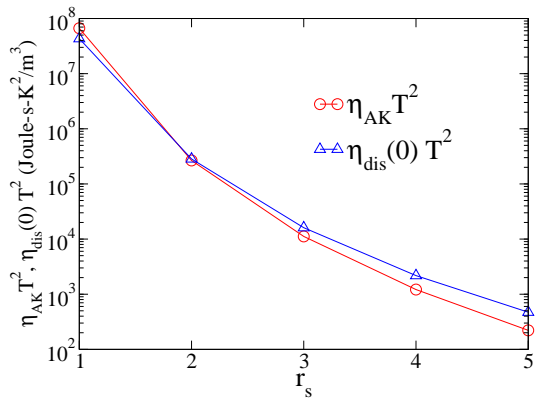}
\end{center}
\caption{The comparison between the viscosity calculated by Abrikosov and Khalatnikov,
$\eta_{AK}$, and the viscosity calculated using the dispersion relation~(\ref{Maxwell}) for different values of $r_s$. The connecting lines are a guide to the eye.}
\label{viscosityAK}
\end{figure}

As a first test of the reasonableness of this approach and also to lend support to its quantitative accuracy, let us apply it to the calculation of $\eta(0)$ for the homogeneous electron liquid, with $\mu_{\infty}$ the {\it full} high-frequency shear modulus given, for example, in Eq. (103) of Ref.~\cite{Tokatly05}.  This can be written as
\begin{eqnarray}
\mu_{\infty}&=&\mu_s+\mu_{xc,\infty}\nonumber \\
&=&\frac{2}{5}n\epsilon_F-n\left[\frac{14}{15}\epsilon_{xc}(r_s)+\frac{4}{5}r_s\epsilon_{xc}'(r_s)\right],
\label{muxc}
\end{eqnarray}
where $\mu_{s}=2n\epsilon_F/5$ is the {\it non-interacting} shear modulus (arising from Pauli exclusion principle) with $\epsilon_F$ being the Fermi energy, and the remainder, $\mu_{xc,\infty}$, is expressed in terms of the exchange-correlation energy per particle, $\epsilon_{xc}$, and its derivative $\epsilon_{xc}'$ with respect to the electron liquid parameter $r_s$. \cite{Perdew92}
In the homogeneous electron liquid the momentum relaxation time, $\tau$, is limited only by electron-electron interactions and therefore diverges with temperature $T$ as $1/T^2$:\cite{Giuliani05}
\bea
\f{1}{\tau_{ee}}=\f{\pi}{8\hbar\epsilon_F}\f{(\varepsilon_{\bf k}-\epsilon_F)^2+(\pi k_B T)^2}{1+e^{-(\varepsilon_{\bf k}-\epsilon_F)/k_BT}}\xi_3(r_s)\label{tauee}
\eea
where $\varepsilon_{\bf k}$ is the energy of the quasi-particle. For a typical density of $r_s=3$, $\xi_3(r_s=3)\simeq 0.5$ and $\tau_{ee}\simeq 5\times 10^{-7}/T^2$ s at the Fermi energy. For this
set of parameters, from Eq.~(\ref{Maxwell}) we then obtain $\eta(0)=\mu_{\infty}\tau_{ee}\sim 1.6 \times10^4/T^2$ Joule-s/m$^3$ which is in good agreement with $\eta_{AK} =1.12 \times10^4/T^2$ Joule-s/m$^3$. The same level of agreement is found for a wide range of $r_s$ values of typical metallic systems, as shown in Fig.~\ref{viscosityAK}. It is also worth stressing that the agreement we find with our dispersion
relation approach is extremely good for $r_s<4$ despite the assumptions we have made in reaching Eq.~(\ref{Maxwell}). The reason why the agreement between the two viscosities decreases with increasing $r_s$
can be attributed to the fact that Eq.~(\ref{tauee}) for the relaxation time is strictly valid for
small values of $r_s$. These results thus lend strong support to the present approach to compute viscosity in terms of the relaxation time $\tau$.

\section{An integrable model}
\begin{figure}[t]
\begin{center}
\includegraphics[width=5cm]{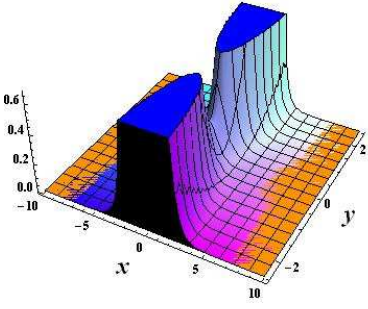}
\end{center}
\caption{A schematic of the potential~(\ref{pot}) - for a constant value of $z$ - used in this work to represent the scattering off a nanojunction.}
\label{model}
\end{figure}

Having checked that the dispersion relation~(\ref{Maxwell}) gives both a qualitative and a quantitative account of the d.c. viscosity we can now proceed to the estimate of the latter in the presence of a nanojunction. We consider the following confining potential (see Fig.~\ref{model} for its schematic) which - despite its simplicity - is a reasonable choice to mimic the scattering properties of nanostructures
\bea
V(x,y,z)=\f{A+B(y^2+z^2)}{\cosh^2(\alpha x)}~.\label{pot}
\eea
where $x$ is along the direction of charge flow. The parameters $A$ and $B$ determine the strength of the potential in the longitudinal and transverse direction, respectively, and $1/\alpha$ measures the extension of the potential along the $x$ direction. Here, we are interested to study transport on a length scale such that $1/\alpha\ll l_{ee}~\simeq \sqrt{D\tau_{ee}}$, with $D$ being the diffusion constant. This simple integrable model allows us to calculate both the elastic momentum relaxation time $\tau_{\bf k}$, the longitudinal transmission coefficient $T_{k_x}$ (defined as the transmission probability for processes which do not change the incident transverse momentum), and the equilibrium density $n(x)$. In the linear-response regime, $\tau_{\bf k}$ due to boundary scattering from the potential~(\ref{pot}) can be estimated by employing the generalized Fermi golden rule,
\bea
\f{1}{\tau_{\bf k}}=L_xL_{\perp}^2\int_{{\bf k}'\ne {\bf k}} \f{d{\bf k}'}{4\pi^2\hbar}\delta(\epsilon_{{\bf k}'}-\epsilon_{\bf k})~|\la \phi_{\bf k}|V(x,y,z)|\psi_{{\bf k}'}\ra|^2,
\label{momrelax2}
\eea
with $L_x \sim l_{ee}$ along $x$ and $L_{\perp}$ being the length of the electrodes in the direction perpendicular to $x$. We use $\epsilon_{\bf k}=\hbar^2k^2/2m_e$ with $m_e$ being the electron mass. The generalized Fermi golden rule allows us to study the full range of longitudinal transmission  for all values $V(x,y,z)$, while in the Born approximation of the Fermi golden rule one could only study longitudinal transmission near unity since its validity is limited for $V(x,y,z)\ll E_{F}$.

We choose the incident wave-function $\phi_{\bf k}(x,y,z)$ with three components of initial momentum $k_x,k^j_y,k^{j'}_z$ as a combination of plane-waves with proper boundary conditions, i.e., open boundary condition along the direction of transport and $\phi_{\bf k}(x,y=0,L_{\perp},z)=\phi_{\bf k}(x,y,z=0,L_{\perp})=0$.
\bea
\phi_{\bf k}(x,y,z)=\f{2}{\sqrt{L_x}L_{\perp}}~e^{ik_x x}\cos(k^j_y y)\cos(k^{j'}_z z)~,\label{instate}
\eea
We use the exact scattering state $\varphi_{k'_x}(x)$ of the one-dimensional potential $1/\cosh^2(\alpha x)$ along the $x-$direction and plane waves with the same boundary conditions as the incident state in the transverse directions. For the transverse wave-functions, plane-waves are a better choice than the harmonic potential eigenstates in the asymptotic region,
since in that region the curvature of the harmonic potential becomes infinitesimally small. Surely, the use of plane waves - instead of an exact scattering state $|\psi_{{\bf k}'}\ra$ of the full potential in Eq.~(\ref{pot}) - would seem a limitation of our calculations. However, we stress that the use of an asymptotic state of $\psi_{{\bf k}'}(x,y,z)$ is also not the right choice in Eq.~(\ref{momrelax2}),
since the asymptotic state is a better approximation only far away from the action of the potential (where $V(x,y,z)=0$), while $|\la \phi_{\bf k}|V(x,y,z)|\psi_{{\bf k}'}\ra|$ of Eq.~(\ref{momrelax2}) is nonzero only for $V(x,y,z) \ne 0$.

The scattered wave-function $\psi_{{\bf k}'}(x,y,z)$ with momentum $k'_x,k^p_y,k^{p'}_z$ is thus
\bea
\psi_{{\bf k}'}(x,y,z)=\f{2}{\sqrt{\mathcal{N}}L_{\perp}}\varphi_{k'_x}(x)\cos(k^p_y y)\cos(k^{p'}_z z)~, \label{scattstate}
\eea
with
\begin{widetext}
\bea
\varphi_{k'_x}(x)=(1-\xi^2)^{-ik'_x/2\alpha}{\rm F}[-ik'_x/\alpha-s,-ik'_x/\alpha+s+1,-ik'_x/\alpha+1,(1-\xi)/2]~,
\eea
\end{widetext}
and $\xi={\rm tanh}(\alpha x)$, where $k^p_y=(2p+1)\pi/L_{\perp}$ and $k^{p'}_z=(2p'+1)\pi/L_{\perp}$with $p,p'=0,\pm1,\pm2..$, $s=(-1+\sqrt{1-8m_eA/\alpha^2\hbar^2})/2$ and  $F(\beta,\gamma,\delta,z)$ is the hypergeometric function. The normalization factor $\mathcal{N}$ is fixed by $\int_{-L_x/2}^{L_x/2}dx~|\varphi_{k'_x}(x)|^2=\mathcal{N}$.

We separately evaluate $\tau^{||}_{\bf k}$, $\tau^{\perp}_{\bf k}$ the longitudinal and transverse relaxation times, respectively, with $1/\tau_{\bf k}=1/\tau^{||}_{\bf k}+1/\tau^{\perp}_{\bf k}$. We include in $\tau^{||}_{\bf k}$ only those contributions of momentum relaxation which do not change the index of incident transverse channels, while all other processes are included in $\tau^{\perp}_{\bf k}$.
Due to its symmetry in the transverse plane, the potential in Eq.~(\ref{pot}) does not allow processes in which the $y$ and $z$ components of the incident momentum change simultaneously. For example, if we choose the incident state with $j,j'=0$, the allowed contributions in $\tau^{\perp}_{\bf k}$ come from the channels with indices $p=1,2,\cdots,p_c$, $p'=0$ and $p=0$, $p'=1,2,\cdots,p_c$, where $p_c$ is the maximum number of allowed transverse channels.
\begin{figure}[t]
\includegraphics[width=7cm]{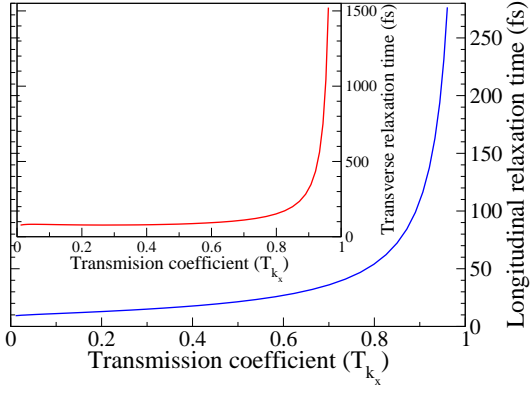}
\caption{Longitudinal momentum relaxation time $\tau^{||}_{\bf k}$ as a function of the transmission coefficient $T_{k_x}$ along the $x$-direction for $r_s=3$. Inset shows the corresponding transverse momentum relaxation time $\tau^{\perp}_{\bf k}$.}
\label{relaxee}
\end{figure}

On physical grounds of local thermodynamic equilibrium in the electrodes, we can assume that the incident state is prepared with the lowest transverse momentum component, i.e., $j,j'=0$. Therefore, for an incident state at the Fermi energy, $k_x=k_F\sqrt{1-2(\pi/k_FL_{\perp})^2}$ and $k^0_y=k^0_z=\pi/L_{\perp}$. The longitudinal transmission coefficient $T_{k_x}$ along $x$ for  the longitudinal momentum $k_x$ of the incident state is calculated from $\varphi_{k_x}(x)$, and is given by \cite{Landau97}
\bea
T_{k_x}=\f{\sinh^2(\pi k_x/\alpha)}{\sinh^2(\pi k_x/\alpha)+\cosh^2(\f{\pi}{2}\sqrt{8m_eA/\hbar^2\alpha^2-1})}~,
\eea
with $8m_eA/\alpha^2\hbar^2>1$.
We find after some algebra
\bea
\f{1}{\tau^{||}_{\bf k}}&=&\f{m_e}{\hbar^3|k_x|}|I(-k^0_x)|^2~\big(A+BL_{\perp}^2\f{(\pi^2-6)}{6\pi^2}\big)^2~,\label{longrelax}\\
\f{1}{\tau^{\perp}_{\bf k}}&=&\f{m_eB^2L_{\perp}^4}{\pi^4\hbar^3}\sum_{s=\pm 1,p=1,2..}^{p_c}\f{1}{|k^p_x|}\f{(2p+1)^2}{p^4(p+1)^4}|I(sk^p_x)|^2,~~\label{tranrelax3}
\eea
where $I(k^p_x)=(1/\sqrt{\mathcal{N}})\int_{-L_x/2}^{L_x/2}dxe^{-ik_xx}\varphi_{k^p_x}(x)/\cosh^2(\alpha x)$.  $k^p_x~[=k_F\sqrt{1-((2p+1)^2+1)\pi^2/(k_FL_{\perp})^2}]$ is the longitudinal momentum of the scattered electron in the transverse channel of indices $p$ and $p'=0$.

The transverse and longitudinal relaxation times are plotted in Fig.~\ref{relaxee} for the electron density $r_s=3$, and typical dimensions of nanoscale junctions, namely $L_{x}=20$~nm, $L_{\perp}=5$~nm and $1/\alpha=2$~nm. We always keep the transverse confinement potential strength ($BL_{\perp}^2>55$ eV) at far above the Fermi energy ($\epsilon_F=5.57$ eV for $r_s=3$), and tune the height ($A$) of the longitudinal barrier across the Fermi energy to have different values of the transmission coefficient $T_{k_x}$.

Figure~\ref{relaxee} confirms our initial hypothesis, namely that the elastic relaxation time is the dominant contribution to the total relaxation time for $|{\bf k}|=k_F$ and for a wide range of temperatures:
$1/\tau=1/\tau_{ee}+1/\tau_{\bf k}\approx 1/\tau_{\bf k}$. We finally note that if we start with an incoming state with $j,j'\ne 0$, then the incident longitudinal momentum $k_x$ at the Fermi energy is smaller than that of the incident state $j,j'=0$ we have considered so far. The corresponding longitudinal transmission $T_{k_x}$ is also relatively smaller for the same longitudinal barrier height $A$. Thus, we find a faster longitudinal relaxation and a slower transverse relaxation time at the same barrier height. However, if we plot the total relaxation time as a function of the longitudinal transmission coefficient by reducing the barrier height, we find that the total relaxation time is qualitatively and quantitatively similar as for the $j,j'=0$ initial state.

 \begin{figure}[t]
\begin{center}
\includegraphics[width=7cm]{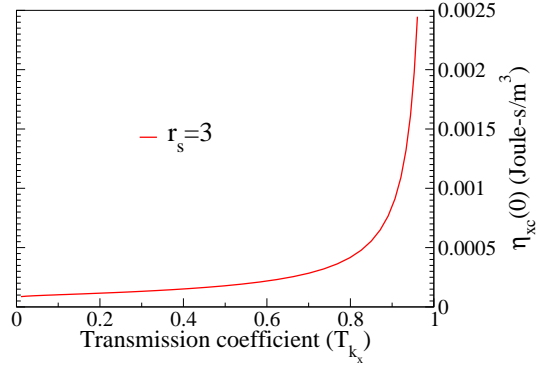}
\end{center}
\caption{Zero-frequency xc viscosity $\eta_{xc}$ in a nanojunction as a function of the transmission coefficient for $r_s=3$.}
\label{viscosity}
\end{figure}

We are now ready to estimate the desired viscosity from Eq.~(\ref{Maxwell}), and from Eq.~(\ref{dynres}) the dynamical corrections to the single-particle resistance. For this we need only the xc component of the shear modulus, since the xc viscosity in the d.c. limit is $\eta_{xc}(0)=\mu_{xc,\infty}\tau_{F}$. This quantity is plotted in Fig.~\ref{viscosity}  with $\mu_{xc,\infty}$ given by the second term on the right hand side of Eq.~(\ref{muxc}) and $\tau_{F}$ evaluated above for different values of $T_{k_x}$.

\begin{figure}[t]
\begin{center}
\includegraphics[width=7cm]{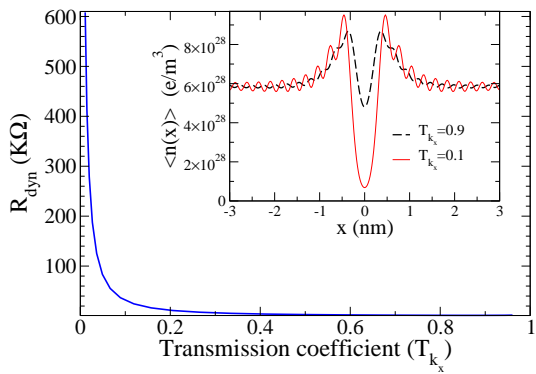}
\end{center}
\caption{Dynamical viscous resistance as a function of transmission coefficient for the potential~(\ref{pot}). The inset shows the averaged density $\la n(x)\ra$ across the nano-junction for the longitudinal transmission coefficient 0.1 and 0.9. Other parameters are in the text.}
\label{DynRest2}
\end{figure}

Finally, since in the present model the transverse variation of the density is small, we evaluate only the longitudinal contribution to $R_{dyn}$ from Eq.~(\ref{dynres}) using the full wavefunction $\psi_{{\bf k}}(x,y,z)$ and by
defining the planar average
\bea
\la n(x) \ra=\f{1}{L_{\perp}^2}\int_{-L_{\perp}/2}^{L_{\perp}/2} \int_{-L_{\perp}/2}^{L_{\perp}/2}dy dz \sum_{{\bf k}\le{\bf k}_F}|\psi_{{\bf k}}(x,y,z)|^2
\eea
with the proper normalization of $\la n(x) \ra$ corresponding to the bulk density of $r_s=3$ deep into the leads. This quantity is plotted in Fig.~\ref{DynRest2}. Introduction of the potential $V(x,y,z)$ induces two effects: one is the reduction of the xc viscosity entering Eq.~(\ref{dynres}) due to fast momentum relaxation, and the other is the increase of the density gradient. These two effects together determine the value of the dynamical resistance corrections in Fig.~\ref{DynRest2}. In agreement with what was previously
reported~\cite{Sai05}, this dynamical resistance is relatively small at large transmissions due to
the small variation of the density across the junction. However, it increases substantially at very low transmissions with values that can greatly exceed, for this particular model, the dynamical resistances estimated previously~\cite{Jung07}.

\section{Conclusion}
In conclusion, we have introduced a dispersion-relation approach to estimate the d.c. viscosity which provides very good agreement with the estimates obtained using the standard
non-perturbative calculation for the homogeneous electron liquid. This approach tremendously simplifies the calculations of the viscosity and allows to estimate this quantity in the presence of a nano-constriction where
the momentum relaxation time is dominated by the elastic collisions at the junction. We have then computed the many-body contribution to the resistance of the junction for an integrable potential and found that while this resistance
is relatively small for transparent barriers it is substantially higher for low transmission barriers, a fact which goes in the right direction in explaining the well-known (and yet unsolved) discrepancy between theory and experiments in molecular junctions \cite{MDlong}.

\section{Acknowledgement}
DR and MD acknowledge support from the DOE grant DE-FG02-05ER46204 and UC Laboratories, GV from DOE under Grant No. DE-FG02-05ER46203.

\end{document}